\documentclass{emulateapj}

\shorttitle{Interferometry on the hot corino of IRAS16293}
\shortauthors{Bottinelli et al.}

\newcommand{\kms}          {\mbox{${\rm km~s^{-1}}$}}
\newcommand{\jyb}          {\mbox{${\rm Jy~beam^{-1}}$}}
\newcommand{\vlsr}          {\mbox{$V_{\rm LSR}$}}

\def\ch3cn{\mbox{CH$_3$CN}}
\def\hcooch3{\mbox{HCOOCH$_3$}}

\tabletypesize{\scriptsize}

\begin{document}

\title{Near-arcsecond resolution observations of the hot corino of 
the solar type protostar IRAS 16293--2422$^0$
}

\author{S. Bottinelli\altaffilmark{1,2}, C. Ceccarelli\altaffilmark{1}, 
R. Neri\altaffilmark{3}, J. P. Williams\altaffilmark{2},
E. Caux\altaffilmark{4}, S. Cazaux\altaffilmark{5},
B. Lefloch\altaffilmark{1},
S. Maret\altaffilmark{1}, A. G. G. M. Tielens\altaffilmark{6}}

\altaffiltext{0}{Based on observations carried out with the IRAM 
Plateau de Bure Interferometer. IRAM is supported by INSU/CNRS (France), 
MPG (Germany) and IGN (Spain).}
\altaffiltext{1}{Laboratoire d'Astrophysique de l'Observatoire de Grenoble, 
BP 53, 38041 Grenoble, Cedex 9, France. \\
sbottine@obs.ujf-grenoble.fr; 
ceccarel@obs.ujf-grenoble.fr;\\
lefloch@obs.ujf-grenoble.fr;
maret@obs.ujf-grenoble.fr}
\altaffiltext{2}{Institute for Astronomy, University of Hawai`i, 
2680 Woodlawn Drive, Honolulu HI 96822, USA.
jpw@ifa.hawaii.edu}
\altaffiltext{3}{Institut de Radio Astronomie Millim\'etrique, 300 rue de 
la Piscine, 38406 Saint Martin d'H\`eres, France. neri@iram.fr}
\altaffiltext{4}{Centre d'Etude Spatiale des Rayonnements, CNRS-UPS, 9 Avenue
du Colonel Roche, BP 4346, 31028 Toulouse, Cedex 4, France.\\
caux@cesr.fr}
\altaffiltext{5}{INAF, Osservatorio Astrofisico di Arcetri, 
Largo Enrico Fermi, 5
I-50125 Firenze, Italy. cazaux@arcetri.astro.it}
\altaffiltext{6}{Kapteyn Astronomical Institute, P.O. Box 800, 9700 
AV Groningen, The Netherlands. tielens@astro.rug.nl}

\begin{abstract}
Complex organic molecules have previously been discovered in solar type 
protostars, raising the questions of where and how they form in the envelope. 
Possible formation mechanisms include grain mantle evaporation, interaction
of the outflow with its surroundings or the impact of UV/X-rays inside
the cavities.
In this Letter we present the first interferometric observations 
of two complex molecules,
CH$_3$CN and HCOOCH$_3$, towards the solar type protostar IRAS16293--2422.
The images 
show that the emission originates from two compact regions centered on the
two components of the binary system. We discuss 
how these results favor the grain mantle evaporation scenario and
we investigate the implications of these observations for the chemical
composition and physical and dynamical state of the two components.
\end{abstract}


\section{Introduction}

Solar type Class 0 protostars are characterized by being observable only
at millimeter to far infrared wavelengths.
This is because the central forming stars are surrounded and obscured 
by massive envelopes, making them the coldest known protostars.
Yet, the cold envelopes hide inner warm regions where the grain mantles, built
during the pre-collapse phase, evaporate. 
Following gas phase reactions between the evaporated mantle molecules
(also called first generation molecules), 
organic complex molecules (second generation molecules) form in these regions.
While this picture has now been widely accepted for massive protostars
(see e.g. Kurtz et al. 2000; van Dishoeck \& Blake 1998), it is 
only recently that evidences in support of parts of this scenario 
have been accumulated for low mass protostars.
Actually, so far the existence of inner warm regions with evaporated mantles
in solar type protostars has been argued indirectly from the sophisticated 
analysis of molecular multi-frequency single dish observations 
(Ceccarelli et al. 2000a,b; Sch\"oier et al. 2002; Maret et al. 
2002, 2004; Doty et al. 2004).
A survey on almost a dozen of Class 0 sources has shown that, typically, 
the inner warm regions have sizes of a few tens of AUs  
(Maret et al. 2004), i.e the size of the Solar Nebula.
Now that the existence of these regions has been demonstrated,
the questions of their nature and chemical composition
have become relevant because of the link with the formation of our own
Solar System. 
What is the inventory of organic molecules of such regions?
What are their origin and evolution?
In particular, could these molecules be incorporated into the planet-forming
disks surrounding the protostars?

Based on theoretical arguments, no complex, second generation molecules
should have the time to form in solar type protostars. The reason is that 
the gas crossing time
of the warm regions is estimated to be much shorter ($\lesssim 10^3$ yr), 
than the predicted chemical formation time ($\gtrsim 10^4$ yr;
Sch\"oier et al. 2002).
But our theories have somewhere flaws, 
since complex organic molecules have been detected
in the two (out of two) sources where they have been 
searched for: IRAS16293--2422 (Cazaux et
al. 2003) and NGC1333-IRAS4A (Bottinelli et al. 2004).
We therefore may even doubt the basic prediction that those molecules
originate in the warm inner regions of the envelopes!
It is, for example, possible that the process which is responsible for the 
release in the gas phase of the mantle constituents,
is not the thermal evaporation, but the
interaction of the outflow with the surroundings, or even of UV/X-rays with
the cavities excavated by the outflow (Sch\"oier et al. 2002, 2004).
Alternatively, the emission could come from disk surfaces which would 
have been heated by accretion shocks (Sch\"oier et al. 2004).
The proof that the complex organic molecules detected by Cazaux et al. (2003)
and Bottinelli et al. (2004) originate from the inner warm regions
(envelope or disk), 
called hot corinos (Ceccarelli 2004; Bottinelli et al. 2004), 
was still missing until today.

In this Letter we bring the missing proof and
report interferometric observations of two complex,
second generation molecules, methyl cyanide and methyl formate, towards
the prototype of the hot corinos, IRAS16293--2422 (hereafter IRAS16293).

\section{Observations and results}

Observations of IRAS16293 
($\alpha$(2000) = $16^{\rm h}32^{\rm m}22\fs 6$, 
$\delta$(2000) = $-24^{\circ}28'33''$) were carried out at the IRAM Plateau
de Bure Interferometer on February 1$^{\rm st}$ and March 25$^{\rm th}$ 2004 
in the B and C configurations of the array. 
Five \ch3cn transitions at 110.4 GHz and
4 \hcooch3 transitions at 226.4 GHz were obtained simultaneously, along with 
the continuum emission at 3~mm and 1.3~mm. The receivers were tuned single side
band at 3~mm and double side band at 1.3~mm. \ch3cn and \hcooch3 transitions
were covered with two correlator units, each of 40 and 80 MHz bandwidth 
respectively. Typical system temperature were 250~K (USB) at
3~mm and 500~K (DSB) at 1.3~mm. Phase and amplitude calibrations were obtained
by observing the nearby point sources 1514-241 and
NRAO 530 every 20 minutes.
Bandpass calibration was carried out on 3C273 and 0851+202 and the 
absolute flux density scale was derived from MWC349, 3C345 and 0923+392.
Data calibration was performed in the antenna based manner and uncertainties
are less than 10\% at 3~mm and less than 20\% at 1.3~mm. 
Flux densities were obtained from visibilities using standard IRAM
procedures. Continuum images were produced by averaging line-free channels.
Line maps were obtained by cleaning line images after subtraction of 
the continuum directly from the visibilities. 

\begin{figure*}
\centering
\includegraphics[angle=270,width=7.9cm]{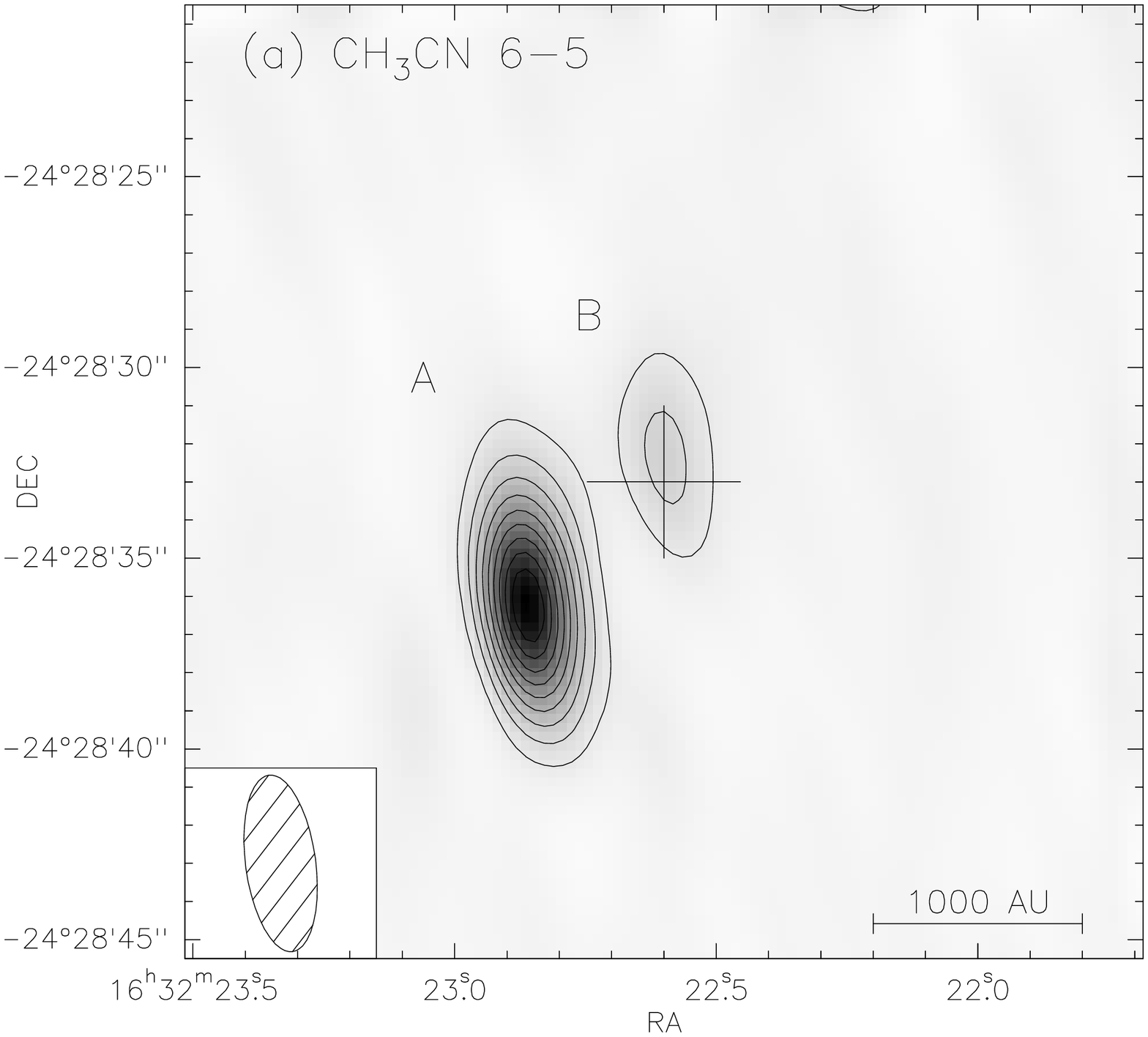}
\includegraphics[angle=270,width=7.9cm]{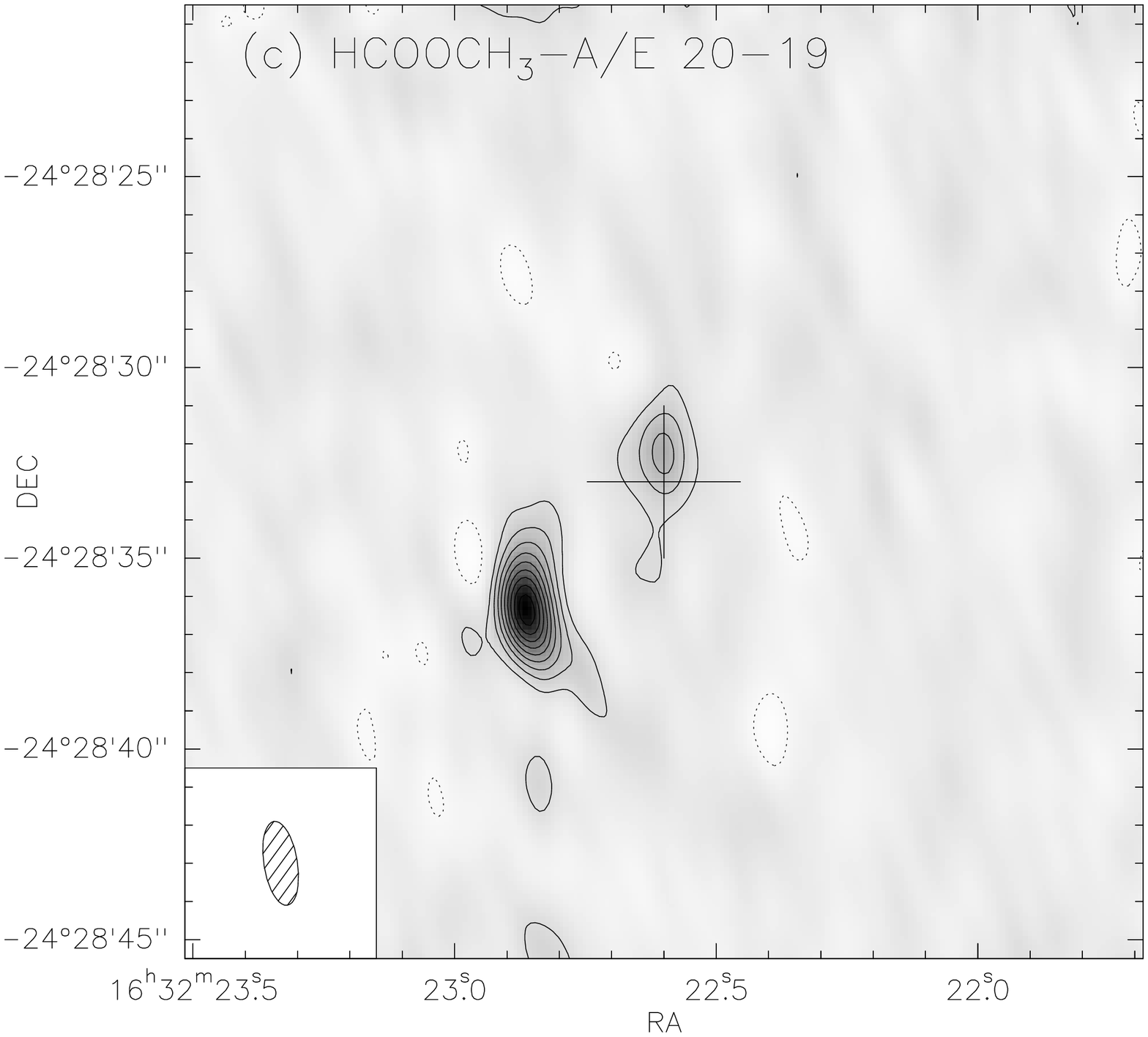}\\
\includegraphics[angle=270,width=7.9cm]{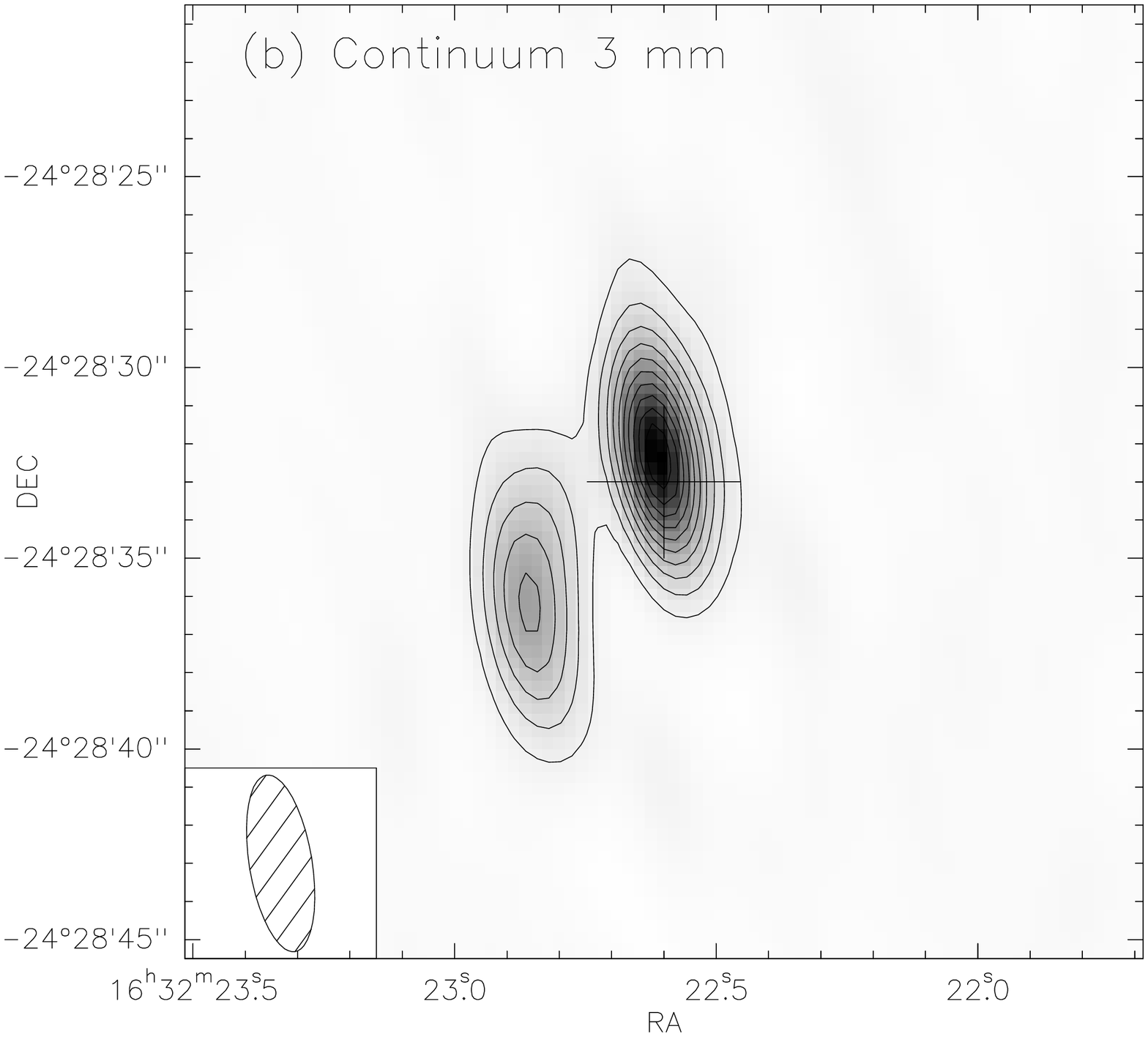}
\includegraphics[angle=270,width=7.9cm]{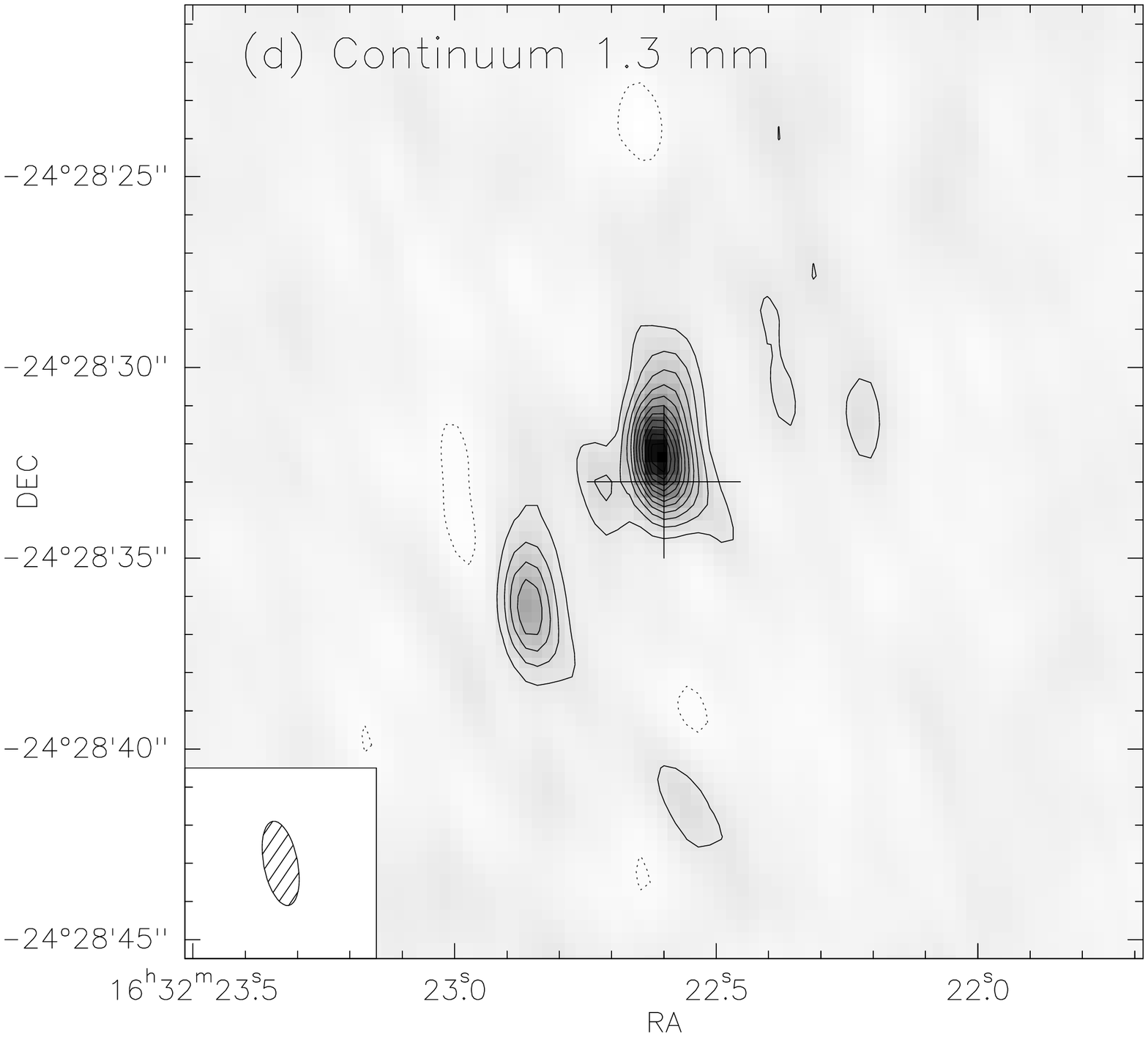}\\
\caption{(a) Line map of \ch3cn, averaged over the five transitions
listed in Table 1; the rms is 3 m\jyb and 
contours range from 15 to 150 m\jyb\ in steps of 15 m\jyb.
(b) Continuum emission at 3~mm, with an rms of 3 m\jyb; 
contour levels range from 20 to 220 m\jyb\ in steps of 20 m\jyb.
(c) Line map of \hcooch3-A and -E, averaged over the four transitions
listed in Table 1; the rms is 8 m\jyb and 
contour levels range from 20 to 200 m\jyb\ in steps of 20 m\jyb.
(d) Continuum emission at 1.3~mm, with an rms of 15 m\jyb; 
contour levels range from 50 to 600 m\jyb\ in steps of 50 m\jyb.
Beam sizes are 
$4\farcs 7\times1\farcs 6$ and $2\farcs 2\times0\farcs 9$ 
at 3 and 1.3~mm respectively.}
\label{linemaps}
\end{figure*}

Figures~\ref{linemaps}-a and \ref{linemaps}-c show the integrated line emission
of \ch3cn and \hcooch3, averaged over all the transitions listed in Table 1 for
each molecule. Note that the energy of the upper level of the \ch3cn transitions
decreases with frequency, but that maps averaged over each individual 
transition do not show any significant difference. This means that 
the emitting region does not depend
on the energy of the transition, i.e. on the excitation conditions (but rather on a jump
of the molecular abundances).
Continuum emission at
3 and 1.3~mm is displayed in Figures~\ref{linemaps}-b and \ref{linemaps}-d
respectively . These maps show two components 
which are spatially coincident with the centimeter wavelength emission regions
A and B mapped by Wooten (1989) and with the millimeter wavelength 
emission regions MM1 and MM2 mapped by Mundy et al. (1990, 1992). 
As already noted in the previously mentioned
works (see also Looney, Mundy \& Welch 2000; Sch\"oier et al. 2004), 
the south-east region (``source A'' or ``MM1'') is the weakest in the 
continuum but brightest in line emission. 
On the contrary, the north-west region (``source B'' or
``MM2'') is the brightest in the continuum and weakest in line emission.

\begin{deluxetable*}{c|ccc|cc|cc|cc}
\tablewidth{0pt}
\tablecaption{Line and continuum emission}
\tablehead{Molecule & Transition & E$_{\rm up}$\tablenotemark{a} & Frequency (GHz) &
\multicolumn{2}{c|}{(Integrated) intensity} &  
\multicolumn{2}{c|}{Size ($''$)\tablenotemark{b}} &
\multicolumn{2}{c}{$\Delta$RA,$\Delta$Dec ($''$)\tablenotemark{c}}\\
& & (cm$^{-1}$) & or Wavelength & A & B & A & B & A & B}
\startdata
\ch3cn & $6_{5,0}-5_{5,0}$     & 137.1 & 110.330 & 1.39(0.09) & 0.51(0.06) & 0.8(0.2) & $<0.8$ & 3.6,$-3.3$ & $-0.0$,1.1 \\
       & $6_{4,0}-5_{4,0}$     &  92.4 & 110.350 & 1.70(0.08) & 0.38(0.06) & 0.8(0.2) & $<0.8$ & 3.7,$-3.2$ & $-0.1$,0.7 \\
       & $6_{3,0}-5_{3,0}$     &  57.6 & 110.364 & 2.75(0.08) & 0.67(0.05) & 0.9(0.1) & $<0.8$ & 3.7,$-3.2$ & $+0.1$,0.6 \\
       & $6_{2,0}-5_{2,0}$     &  32.8 & 110.375 & 2.56(0.08) & 0.51(0.05) & 1.2(0.1) & $<0.8$ & 3.6,$-3.3$ & $+0.1$,0.9 \\
       & $6_{1/0,0}-5_{1/0,0}$ &  17.9/12.9 & 110.381/110.383 & 5.26(0.10) & 1.05(0.07) & 0.9(0.1) & $<0.8$ & 3.6,$-3.3$ & $-0.0$,0.6 \\
\hline
\hcooch3-E/A & $20_{2,19}-19_{2,18}$ & 83.5 & 226.713/226.718 & 6.44(0.41) & 3.36(0.36) & 1.4(0.2)$\times$0.7(0.1) & 0.8(0.2) & 3.6,$-3.5$ & $-0.1$,0.7 \\
             & $20_{1,19}-19_{1,18}$ & 83.5 & 226.773/226.778 &11.13(0.49) & 4.05(0.39) & 1.6(0.1)$\times$0.8(0.1) & 0.8(0.1) & 3.7,$-3.3$ & $-0.0$,0.7 \\
\hline
continuum & & &   3~mm & 0.17 & 0.26 & 3.4$\times$1.4 & 1.5$\times$0.8 & 3.5,$-2.9$ & +0.2,0.7 \\
continuum & & & 1.3~mm & 0.77 & 1.02 & 3.7$\times$1.2 & 1.5$\times$0.8 & 3.5,$-3.3$ & +0.2,0.6 \\
\enddata
\tablecomments{ The line intensity is in units of
Jy \kms, while the continuum is in Jy. Errors are given in parentheses.}
\tablenotetext{a}{Energy of the upper level of the transition.}
\tablenotetext{b}{FWHM of gaussian fit.}
\tablenotetext{c}{Relative peak position from the fit.}
\end{deluxetable*}

Table 1 gives the intensities and sizes of the line and continuum
emissions.
Within the errors, we recover all the line emission measured 
by the IRAM 30m (Cazaux et al. 2003).

\begin{figure}
\epsscale{0.95}
\plotone{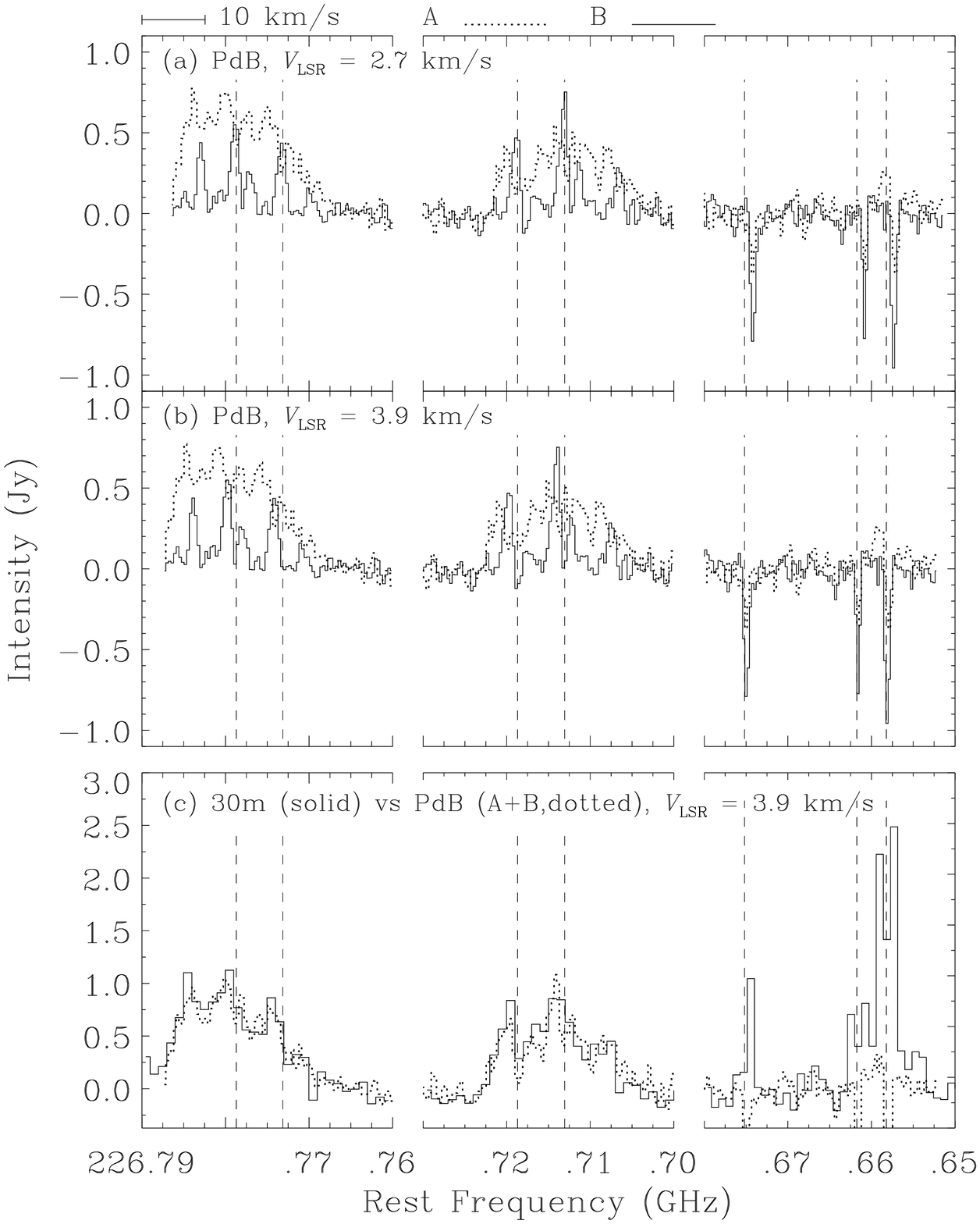}
\caption{(a) Spectra at 1~mm averaged over the emission regions 
of sources A (dotted line) and B (solid line), 
and displayed for \vlsr=2.7 \kms.
The velocity resolution is 0.4 \kms\ and rms are 0.08 and 0.06 Jy
for A and B respectively.
(b) Same as (a) but with \vlsr=3.9 \kms.
(c) The solid line represents the spectrum obtained at the IRAM 30m (Cazaux et al. 
2003) with a velocity resolution of 1.3 \kms and rms of 0.11 Jy.
Overlayed in dotted line is the sum of the Plateau de Bure spectra
of sources A and B (rms=0.10 Jy).
Vertical lines indicate the frequencies
of HCOOCH$_3$ (A and E) lines (emission) and CN lines (absorption).
Right to left are CN $2_{\frac{3}{2},\frac{5}{2}}-1_{\frac{1}{2},\frac{3}{2}}$,
$2_{\frac{3}{2},\frac{1}{2}}-1_{\frac{1}{2},\frac{1}{2}}$,
$2_{\frac{3}{2},\frac{3}{2}}-1_{\frac{1}{2},\frac{1}{2}}$,
\hcooch3-E and -A $20_{2,19}-19_{2,18}$, 
\hcooch3-E and -A $20_{1,19}-19_{1,18}$.
\label{spectra1mm}}
\end{figure}

Finally, the spectra at 1.3~mm of 
emission regions A and B are shown in Figure~\ref{spectra1mm},
assuming a \vlsr=3.9 (Figure~\ref{spectra1mm}-a)
 and 2.7 \kms\ (\ref{spectra1mm}-b)  respectively
(see discussion in the next section).
Again, note that the sum of A and B (Figure~\ref{spectra1mm}-c)
reproduces very well the features in the spectrum obtained at 
the IRAM 30m (Cazaux et al. 2003).

\section{Discussion and Conclusions}

The most important result of the presented observations is 
that the complex molecules observed by Cazaux et al. (2003) originate
in two compact regions, whose diameters are about 1.5$''$ (A) or less (B),
as shown in Figure \ref{linemaps} (and Table 1).
This goes along with the fact that the images do not show any 
evidence of emission associated with the molecular outflows seen at 
larger scales (see below for the discussion on the line profiles).
Also, following the remark in the Introduction, the Plateau de Bure images
do not reveal any evidence of emission from cavities excavated from the
outflows either.
Indeed, the cavities are typically $\sim 20"$ in size, located $\sim 30"$
away from the source along the $^{12}$CO outflow (see e.g. Arce \& Sargent
2004), which would have been easily detected by the Plateau de Bure.
The two regions where the molecular emission comes from, are compact, and while 
source A is barely resolved in the 1mm images\footnote{It is likely that the
emission from source A originates in the disk around this source. A detailed
study of this aspect is postponed to a forthcoming paper.}, source B 
is unresolved.
The measured sizes (1.5$''$ at 160 pc correspond to a radius of
about 120 AU) are remarkably consistent with the
emission coming from a region where the dust temperature exceeds 100 K
in source A (150-200 AU: based on multi-frequency single dish observations: 
Ceccarelli et al. 2000a,b; Sh\"oier et al. 2002), 
and therefore, where the grain mantles evaporate.
Thus, these observations support the basic prediction (from the modeling of 
the single dish observations; Ceccarelli et al. 2000a,b; Cazaux et al. 2003) 
that a hot corino with a radius of about 
150 AU exists inside the cold envelope of IRAS16293, and that in that
region, complex molecules are formed because of grain mantle evaporation.

In addition to that, the Plateau de Bure observations confirm that the two
sources A and B are different, as noted by previous authors (Wootten 1989;
Mundy et al. 1990, 1992). They differ in line intensities and extent
(Fig. \ref{linemaps}), and this corresponds to a difference 
in their chemical composition.
But, before discussing this point, it is necessary to address the 
second most striking
difference in the two sources: their line profiles (Fig. \ref{spectra1mm}).
Source A has clearly broadened spectra (FWHM $\sim 8$ \kms), while
source B shows apparently much narrower profiles (FWHM $\sim 2$ \kms).
Furthermore, the lines of source B seem to peak at \vlsr=2.7 \kms,
whereas the parent cloud velocity is at \vlsr=3.9 \kms\ (compatible with
the spectra of source A, although given the broad profiles
it is difficult to precisely determine the \vlsr\ of source A). 
Note that the cloud's \vlsr\ (3.9 \kms) is very nicely measured
by the two CN absorption lines in the observed band\footnote{The absorption 
originates in the foreground (envelope + cloud) cold gas, which absorbs
the photons emitted in the hot corino regions (the CN emission component, 
being extended, is filtered out by the interferometric observations).}
(Figure \ref{spectra1mm}).
As said, no evidence of outflowing gas is seen in the images, and
also the broad line profiles of source A are consistent with gas {\it infalling}
towards a $\sim$1 M$_\odot$ object (Ceccarelli et al. 2000a,b; Sch\"oier et al.
2002). Therefore, both the images and line profiles of source A
are fully consistent with the hot corino hypothesis.

The case of source B is less obvious: why does this source have narrower
lines and why do they peak at 2.7 \kms ? But is this true?
A very careful look at the B spectra rises doubts. 
Indeed, all the B lines have a second small peak ---sometimes
at the limit of the noise--- on the red-shifted side of the spectrum. 
This second peak could indeed be part of the line itself, which
would  be strongly self-absorbed at \vlsr=3.9 \kms. If this is the case,
the linewidths of source B would be $\sim$ 4-6 \kms\ (Fig.~\ref{spectra1mm}), 
similar to the linewidths measured towards source A.
Note that the blue peak is expected to be brighter than the red peak in the case
of optically thick lines from infalling gas (Leung \&
Brown 1977; Zhou 1992, 1995; Choi et al. 1995), so it would be consistent
with the $\sim 1$ M$_\odot$ hot corino hypothesis of source B too. 
This alternative explanation, optically thick lines in source B, 
is therefore very appealing and worth exploring in some detail.
Using the LTE approximation,
the required column densities for the \ch3cn and \hcooch3 lines 
to be optically thick are $N\sim10^{16}$ cm$^{-2}$ and
$N\sim10^{17}$ cm$^{-2}$ respectively.
These values are about one order of magnitude larger than the column density
derived in B from the emission lines, assuming that the lines are optically
thin, LTE populated and that the emission region fills up the Plateau de
Bure synthesized beam (see below). Considering that the three adopted
assumptions all underestimate the true column density,
it is indeed possible, but not firmly established, that the lines in
source B are optically thick.
Unfortunately, ``physical'' considerations do not help either
to distinguish between
the two possible interpretations, optically thin or thick lines in source B.
In the first case (B has \vlsr=2.7 \kms\ and FWHM $\sim 2$ \kms), 
source B would be less massive than A and would
revolve around it at 1.2 \kms\ (multiplied by the inclination of the orbit),
at a distance of 800 AU, which is fully consistent with 
$M_{\rm A} \sim 1 M_\odot$ (unless the orbit is in the sky plane).
In the second case (B also has \vlsr=3.9 \kms\ and FWHM $\sim$ 4-6 \kms), A and B 
have comparable masses (similar FWHM), but B is more compact.
Future high resolution observations of optically thin lines are hence required
to definitely settle the question.

As said, the nature of source B affects the determination of the
molecular abundances in this source, and hence,
how much the chemical composition of the A and B hot corinos differ.
Using the relation between column density and observed continuum 
flux density, and using a dust opacity $\kappa_\nu$ of 0.8 cm$^2$ g$^{-1}$
at 1.3 mm and a dust temperature $T_{\rm d}$ of 40 K (e.g. Walker et al. 1986),
we derive molecular hydrogen column densities 
from the 1.3~mm continuum emission equal to
$N({\rm H_2,A})=3.5\times10^{24}$~cm$^{-2}$ in A and
$N({\rm H_2,B})=1.7\times10^{25}$~cm$^{-2}$ in B
(consistent with the values of Mundy et al. (1992)\footnote{We postpone the 
detailed analysis of the continuum emission to a forthcoming paper.}).
Using these values and the \ch3cn total column densities derived from
the rotational diagram method ($N_{\rm A}=1.7\times10^{15}$ cm$^{-2}$ and 
$N_{\rm B}=3.9\times10^{14}$ cm$^{-2}$), we get \ch3cn abundances of 
$4.8\times10^{-10}$ and $2.3\times10^{-11}$ for A and B respectively,
i.e. \ch3cn is $\sim$ 20 times more abundant in A than in B.
A rotational diagram could not be drawn for the \hcooch3 transitions
as they have similar upper energy levels, but assuming $T_{\rm rot}\sim$60 K
(Cazaux et al. 2003) we get \hcooch3 column densities of 
$N_{\rm A}=2.6\times10^{16}$ cm$^{-2}$ and 
$N_{\rm B}=8.4\times10^{15}$ cm$^{-2}$, i.e \hcooch3 abundances of
$7.5\times10^{-9}$ and $4.9\times10^{-10}$ for A and B respectively,
a factor 15 difference\footnote{Note that the abundances quoted in Cazaux
et al. differ from those derived here because of the different estimate
of the H$_2$ column density.}.
Note, however, that since B is unresolved in the line emission,
and the lines could be optically 
thick (see above discussion), the molecular abundances quoted for B
may be underestimated by about an order of magnitude
(note also that the region of molecular emission could 
be more compact than the continuum emission region).
Therefore, resolving the problem of the nature of the observed
line emission in source B is crucial, not only to determine the
dynamical state of this source, but also to correctly assess the difference in the
chemical composition of sources A and B.

In summary, these interferometric observations with the Plateau de Bure
show unambiguously that the observed complex organic molecules are emitted
from two compact ($\le 1.5''$) regions, which is consistent with grain
mantle evaporation. No indication was found in support of other formation
mechanisms for these molecules. Two scenarios have been proposed to explain
the observed differences in sources A and B. However, 
the available data are insufficient
to fully understand how and why the two components differ
in their dynamical and chemical state, and more interferometric observations
(e.g. of optically thin lines) are needed in order to solve the issues raised
in this Letter.

\end{document}